\documentclass[9pt,twocolumn,twoside]{opticajnl}
\usepackage{multirow}  % multirow, multicolumn
\usepackage{threeparttable}
\usepackage{soul}
\journal{ol} % Choose journal (ao,jocn,josaa,josab,ol,optica,pr)

\setboolean{shortarticle}{true}
\title{Suppression of patterning effect using IQ modulator for high-speed quantum key distribution systems}

\author{Yuanfei Gao}
\author[1,*]{Zhiliang Yuan}
\affil[1]{Beijing Academy of Quantum Information Sciences, Beijing 100193, China}
\affil[*]{Corresponding author: yuanzl@baqis.ac.cn}

\begin{abstract}
	Quantum key distribution (QKD) is an attractive technology for distributing secret encryption keys among distant users. The decoy-state technique has drastically improved its practicality and performance, and has been widely adopted in commercial systems.
		However, conventional intensity modulators can introduce security side channels in high speed QKD systems because of their non-stationary working points for decoy-state generation. Here, we analyze the transfer function of an IQ modulator and reveal its superiority for stable decoy-state generation, followed by an experimental demonstration.  Thanks to their convenient two-level modulation and inherent high speed, IQ modulators are ideal for use in high-speed decoy-state QKD systems.
\end{abstract}

\setboolean{displaycopyright}{true}

\begin{document}

\maketitle

\section{Introduction}
	Quantum key distribution (QKD) promises information-theoretically secure exchange of encryption keys between two remote users 	over an optical link \cite{BENNETT14}. 
	Its security relies on transmission of its key materials via single photons.  However, bright telecom-wavelength single photon sources have just emerged in laboratory \cite{Nawrath2022}
	and still require cryogenic cooling and a complex optical setup that prohibits commercial applications.
	For practicality, QKD systems usually use robust and readily available lasers as the light source and mitigate the multi-photon security risk by randomly transmitting decoy pulses that differ just in average intensities from the signal pulses~\cite{Lo05,Wang05}. Today, decoy-state technique has been adopted in almost all practical QKD systems, and remains a \textcolor{black}{powerful} tool in developing next-generation QKD protocols, including measurement-device-independent 
	\cite{lo12,Xie22_AsynchronousMDI-QKD,Zeng22_MP,GU20222167} and twin-field  
	\cite{lucamarini18}, for ultra-long distance quantum communications.
	
	Most decoy-state QKD systems adopt three intensity levels with average photon fluxes of $\mu$, $\nu$ and $\omega$ ($\mu>\nu >\omega\simeq$ 0), termed as signal, decoy and vacuum pulses.  
	Precise intensity control is of paramount importance, as a deviation can severely affect the single-photon parameter estimation and even compromise the QKD security entirely.
	In practical QKD systems, they are generated by passing a pulsed laser output through an intensity modulator (IM), typically an off-the-shelf LiNbO$_3$ device based on a Mach-Zehnder interferometer (MZI) structure, to ensure indistinguishability among quantum signals. 
	As illustrated in Fig.~1a,  the work point for the decoy pulses sits at the slope of the voltage response function so any variation in the driving voltage ($\Delta U$) will translate proportionally into intensity fluctuation ($\Delta I_D$).   
	In high speed QKD systems, IM's finite bandwidth can cause a special type of intensity fluctuation that correlates with modulation patterns~\cite{Yoshino18}. This patterning effect seriously threatens the security of QKD, as current decoy-state security analysis assumes independent and identically distributed signals.
		\begin{figure}[ht]
		\includegraphics[width=\columnwidth]{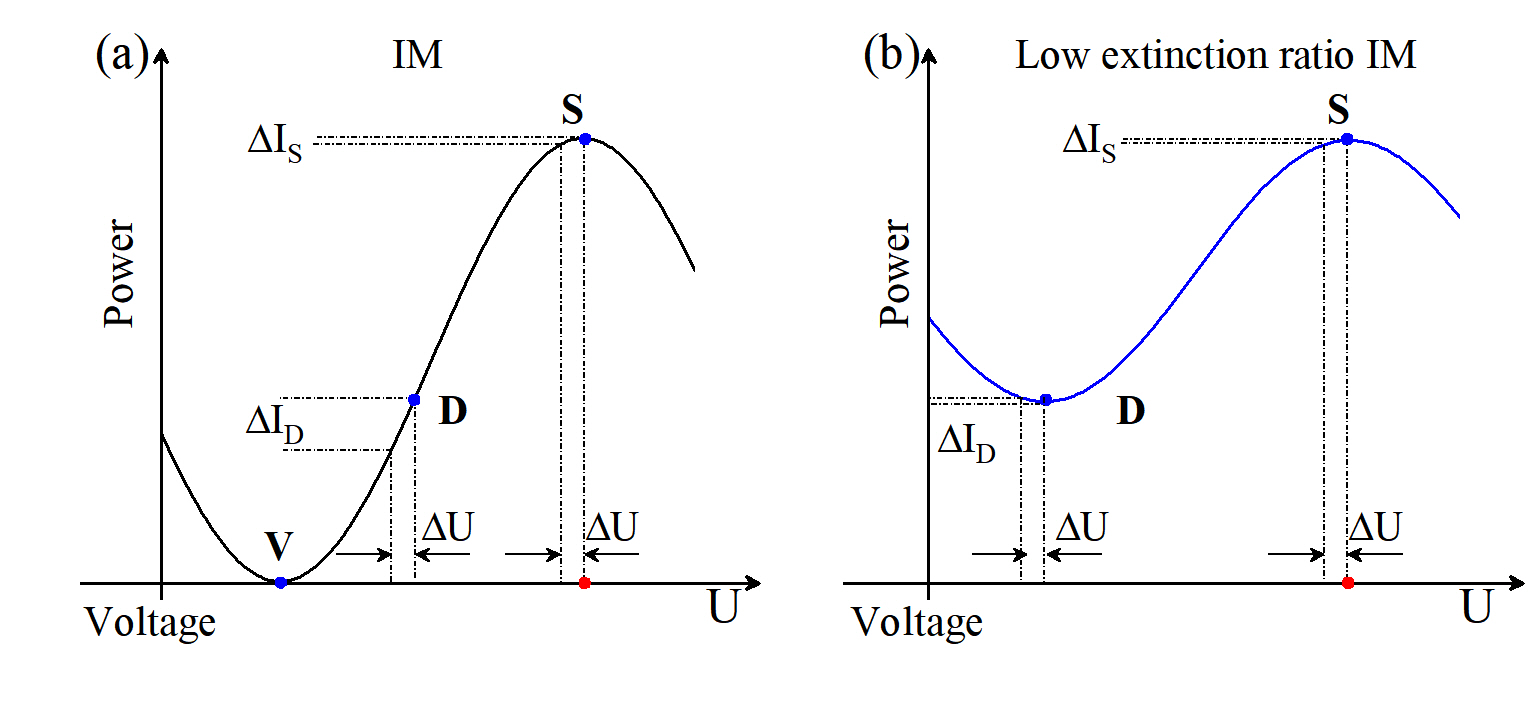}
		\caption{(a) Voltage response function of a commercial LiNbO$_3$-based MZI.  Signal ($\textbf{S}$) and vacuum ($\textbf{V}$) pulses are generated by complete constructive or destructive interference. However, decoy ($\textbf{D}$) pulse is generated at a slope point and can subject to strong intensity fluctuation. 
			(b) Voltage response function of a MZI with non-50/50 beam splitters. 			
		}
	
	\end{figure}
	Approaches to mitigate the patterning effect can be grouped into two broad categories. In software approach, one can either use the worst-case intensity~\cite{wang09} for the single-photon parameter estimation or discard correlated pulses~\cite{Yoshino18} in post-processing, but both will reduce secure key rates significantly.  Alternatively, it is possible to engineer ``imperfect" intensity modulators with low extinction ratios, such as Sagnac\cite{roberts2018} or multi-path MZI\cite{lu2021}, that exploit low visibility interference and thus can produce decoy states also at a stationary modulation point, see Fig.~1b.
	However, neither device is ideal. 
	Sagnac interferometer has an inherent speed limitation due to its time-divisional modulation,
	while a MMZI requires independent intensity and phase control in each of its optical paths and has yet to be realized.  The authors of Ref. \cite{lu2021} simulated a 3-path MZI device with a commercial IQ modulator, but the lack of sufficient path intensity controls may limit
	its capability of producing vacuum states at a stationary modulation point.

	Nevertheless it remains attractive to explore the suitability of IQ modulators for mitigating the patterning effect in decoy-state QKD.  Firstly, IQ modulators have their robustness proven in coherent communication and can therefore be used as a convenient drop-in upgrade to existing QKD systems \cite{RN40}. Secondly, their push-pull driving compatibility allows low driving voltage and chirp-free modulation\cite{RN310}. Thirdly, they can be driven by just binary data lines and can therefore avoid complex multi-level driving electronics that is required for an ordinary IM.  
	In this Letter, we propose and experimentally verify the use of an IQ modulator as a digital 3-level decoy-state IM that can produce a fixed but stable intensity ratio of nominally 1:0.25:0 and can effectively mitigate the patterning effect for decoy-state QKD.  
	As in coherent communication, it needs just two binary driving signals with their amplitudes matching the halfwave voltages of the I and Q modulators.  We numerically analyze the secure key rate (SKR) dependence on the signal to decoy intensity ratio and find the digital IQ modulator is close to optimal.

	\section{Result and discussion}
Figure~2a is a schematic of an IQ modulator, which is constructed with two parallel push-pull Mach-Zehnder modulators nested in a parental MZI. In each of the two modulators, an electrode provides phase shift to two branches, where the shifts are equal in magnitude but with opposite sign. The advantages are that only half the voltage is required as compared with single drive modulation and it does not produce frequency chirp in the optical output. 
We use $\alpha_{I}$, $\alpha_{Q}$ and $\alpha_{G}$ to represent the differential phases of the I-MZI, Q-MZI and IQ-MZI, respectively. $\mu_{\small{in}}$ is the intensity of input pulse. The output of the IQ modulator is written as
\begin{equation}
	\begin{aligned}
		I_{out}
	&= \frac{\mu_{in}}{8}\cdot  \left \{ 2+\cos\left ( \alpha_{\small Q} \right )+ \cos\left ( \alpha_{\small G} \right) + 
	\cos\left ( \alpha_{\small I} \right )\right.\\
	&\left.+		\cos\left ( \alpha_{\small G}+\alpha_{\small I} \right )+
	\cos\left ( \alpha_{\small G}-\alpha_{\small Q} \right )\right.\\
	&\left.	+\cos\left ( \alpha_{\small G}+\alpha_{\small I} -\alpha_{\small Q}\right ) \right \} 		
	\end{aligned}
\end{equation}

Figure~2b shows the output intensity as a function of $\alpha_I$ and $\alpha_Q$.  Here, the $\alpha_{G}$ is set to 0 by adjusting the voltage $DC_{IQ}$. At the point $S\left (\alpha_{I}=0,\alpha_{Q}=0 \right )$, all three modulators undergo constructive interference and we have a maximum intensity output $I_{out} = \mu_{in}$.  This modulation point corresponds to the signal state;
The points $D_{1}\left(\alpha_{I}=0,\alpha_{Q}=\pi\right)$ and $D_{2}\left(\alpha_{I}=\pi,\alpha_{Q}=0\right)$ are saddle points, either of which can be selected to generate the decoy state. The point $V\left (\alpha_{I}=\pi,\alpha_{Q}=\pi \right )$ is a valley point that is used to generate the vacuum state.
The condition of a stationary point is that all its partial derivative must be zero, 
as shown by crossover points in Fig.~2c. Four points \textbf{S}, \textbf{D$_1$}, \textbf{D$_2$} and $V$ correspond to three stationary intensities, and \textbf{D$_1$} and \textbf{D$_2$} are degenerate.

\begin{figure}[ht]
	\includegraphics[width=\columnwidth]{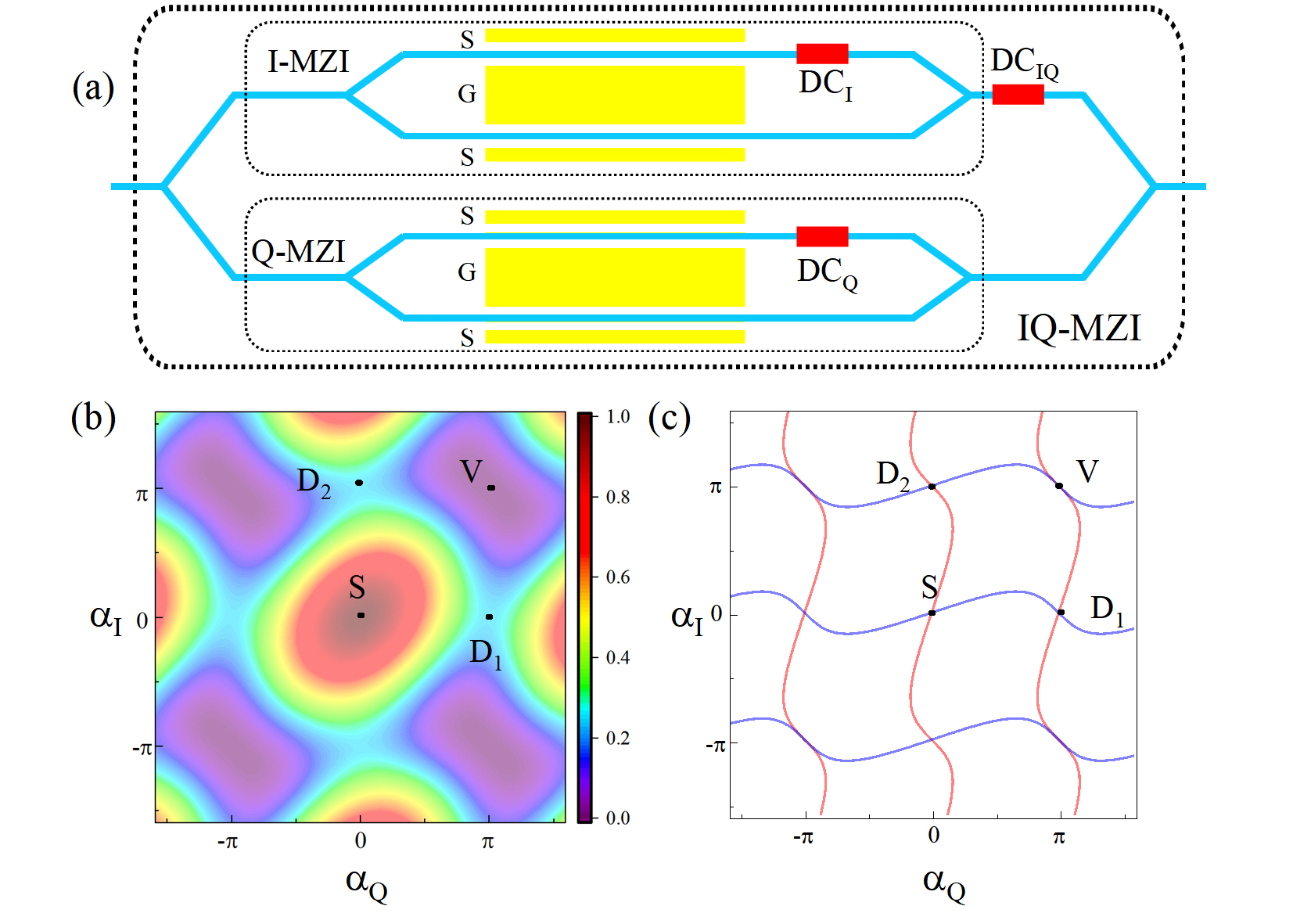}
	\caption{(a)~IQ modulator contains two child push-pull MZIs in parallel. Each MZI can produce a pair of complete constructive and destructive interference. Three stable intensity states can be generated through collaborative work by MZIs.
	(b)~IQ-MZI's output intensity as a function of $\alpha_I$ and $\alpha_Q$.  (c)~Partial derivatives of the intensity output. Crossovers denote stationary points where both partial derivatives equal to zero,  The \textbf{S}, \textbf{D}, and \textbf{V} points are employed as the signal, decoy, and vacuum state respectively.}
	%\label{fig:SKR}
\end{figure}
Whether $I_{MZI}$ and $Q_{MZI}$ are identical determines the number of decoy state intensity generated.
When they are the same , $D_{1}$ and $D_{2}$ have the same intensity $\left(I_{D_{1}}(0,\pi)=I_{D_{2}}(\pi,0)=\frac{\mu_{\small{in}}}{4}\right)$, otherwise, two decoy states are generated. In this case, the encoding of decoy state needs to be fixed as one of them.
Nevertheless, in either case, the points at which the partial derivatives of the IQ modulator are simultaneously zero does not change, and they stay stationary. 
\begin{figure*}[htbp]
	\includegraphics[width=2\columnwidth]{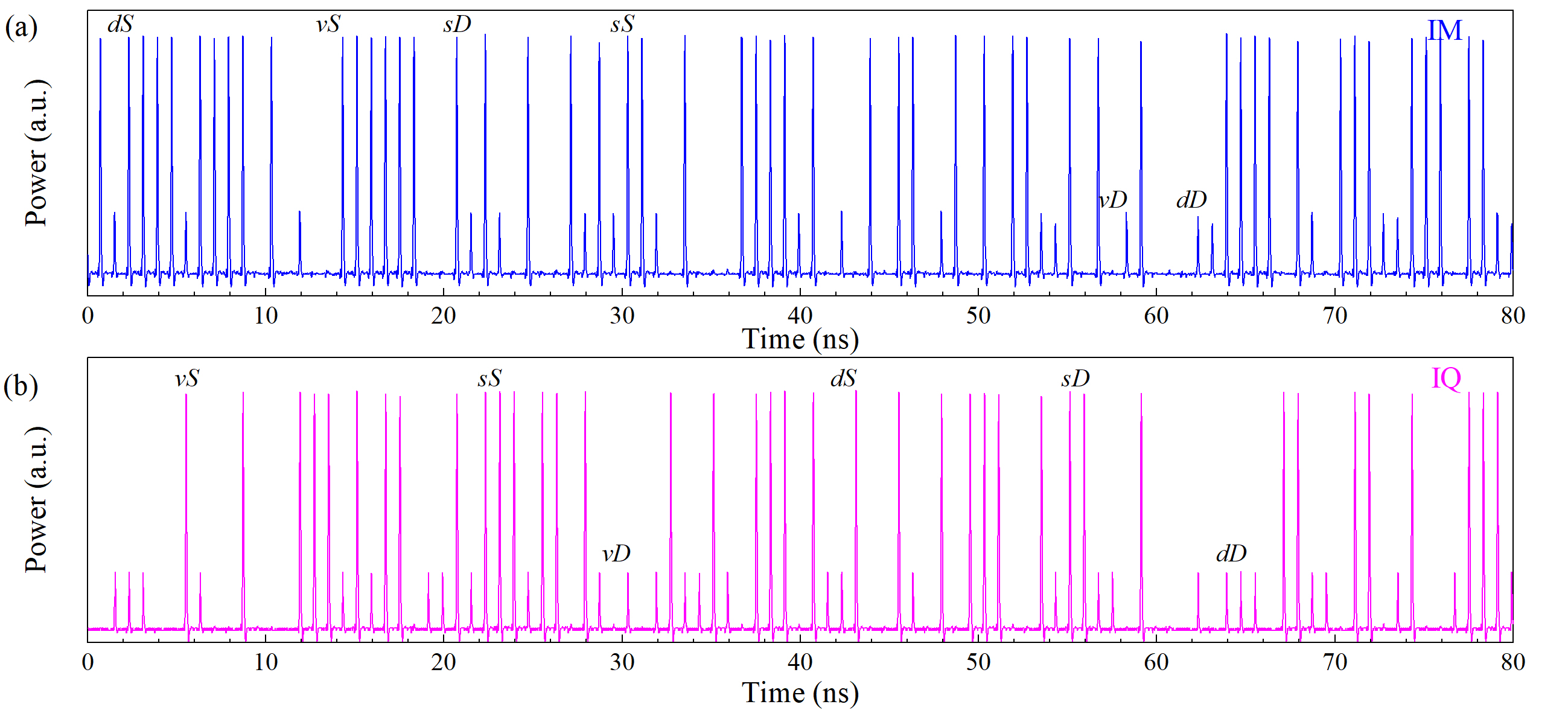}
	\caption{ Oscilloscope traces for IM (a) and IQ (b) modulators including patterns $\textit{sS}$, $\textit{dS}$, $\textit{vS}$, $\textit{sD}$, $\textit{dD}$, $\textit{vD}$.  }
\end{figure*}	

To demonstrate that the IQ modulator can suppress the patterning effect, we characterize the performance of IM and IQ modulator. In our experiments, a 1.25~GHz, 80~ps pulses generated from a distribution feedback laser (DFB) were fed to an IM (Fujitsu, FTM7920FBA) or an IQ (Fujitsu, FTM7961EX) modulators, respectively.  A 25GS/s arbitrary waveform generator with two RF-amplifiers was applied \textcolor{black}{to produce signals of 400~ps width} to drive the modulators. A single three-level modulation signal was used for the IM modulator, while two binary modulation signals were used to drive the IQ modulator.
The pulses are measured by a high-speed photoreceiver with 10 GHz bandwidth and subsequently recorded in an oscilloscope with 16 GHz bandwidth. 
We measured the patterning effect by analysing the intensities of modulated pulses. In the analysis,  the pulse intensity is defined as the area of time profile in one period containing a pulse. 

In the Fig.3, we show the measurements on random patterns to characterize the patterning effect.	Since it is almost localized to adjacent pulses, we consider only adjacent pulses similar to the previous work\cite{Yoshino18}. Here, we also ignore the intensity fluctuation of vacuum state, since its effect on photon detection is smaller than that of stray light and dark counts.
Then, we measure the intensities of $\textbf{S}$ and $\textbf{D}$ pulses with three types of predecessors: 
\textit{s}, \textit{d}, and \textit{v}
%\textbf{S}, \textbf{D}, and \textbf{V} 
pulses. First, we measured the IM intensity modulator. We have accurately controlled the $V_{\pi}$ of the IM modulator to ensure that the \textbf{S} state works at the maximum output.
A short subset of the measured trace is shown in Fig.3(a). When the pulse sequences are $sD$ and $dD$, the difference in intensities of adjacent $\textbf{D}$ states is clearly observable because the IM operates at a non-stationary point for the decoy state. In contrast, 
for the $sS$ and $vS$, the pulse intensity remains almost the same. 

Before measuring the patterning effect for the IQ modulator, we performed a control experiment using identical equipment as used in the above IM characterization.  
Due to the sight difference between the $\textbf{I}$ and the $\textbf{Q}$ modulators, there are two decoy states of different intensities in the output, corresponding to $I_{D1}(0,\pi)$ and  $I_{D2}(\pi,0)$, respectively. 
In order to evaluate the pattern effect, we experimentally choose $D_{1}(0,\pi)$
for the encoding of  \textbf{D} state. 
We also prepared several sequences consisting of the three states shown in Fig.3b.
It can be clearly observed that IQ modulation produces almost identical pulses. The pulse sequence $dD$, $sD$ and $vD$ related to the decoy state maintain almost the same intensity, which is %obviously 
superior to the IM modulator.

To demonstrate that the IQ modulator can be used to eliminate the patterning effect,  we analyze the intensity statistics of different types of pulses recorded under pseudo-random modulations. For each type, \textit{e.g.}, \textit{dD}, we use 10000 pulses to compute its average intensity. 
	Intensities of \textit{sS} and \textit{sD} are used as reference to compute the patterning effect.
Table 1 list the averaged pulse intensities for the different patterns. 
While patterning effects were slight on the $\textbf{S}$ pulses, large deviation about 30 $\%$ was observed on the $\textbf{D}$ pulses. The different behavior of the $\textbf{D}$ pulse comes from the operating point of the decoy pulses. 
At this point, the output intensity is sensitive to the applied voltage fluctuation as shown in Fig.1a.
In contrast, these of the vacuum and signal pulses are set to the extrema of input-output characteristics of the modulator,
so that the output intensities are insensitive to the applied voltage. 
The results indicate that under the same condition,	the patterning effect in IQ modulator is two or three orders of magnitude smaller than the traditional IM modulators.	Employing IQ modulator is highly advantageous to improve %inevitable demand
performance of practical systems.

\begin{center}
\resizebox{\linewidth}{!}{
	\begin{tabular}{ |c|c|c|c|c|c|c| } 
		\hline
		\multicolumn{1}{|c|}{\multirow{2}{*}{Pattern}} & \multicolumn{2}{c|}{IM }&\multicolumn{2}{c|}{IQ }& MMZI  & Sagnac \\ \cline{2-3} \cline{4-5}\cline{6-7}& Avg.int.  &  Patt.eff. & Avg.int.  &Patt.eff.& Patt.eff.& Patt.eff.   \\ \hline
		
		$\textit{sS}$ & 1.000$\pm$ 0.023 & -    & 1.000$\pm$0.034  & - &-&-    \\ 
		\hline
		$\textit{dS}$ & 1.048$\pm$ 0.024 & 4.8$\%$  &1.0001$\pm$0.033 & 0.01$\%$ &-0.07$\%$ &-0.4$\%$   \\ 
		\hline
		$\textit{vS}$ & 1.025$\pm$0.024  & 2.5$\%$  & 0.9996$\pm$0.034 & -0.04$\%$ &0.03$\%$ &-\\
		\hline
		$\textit{sD}$ & 0.241$\pm$0.025  & -    & 0.2529$\pm$0.027 & -&-&-     \\ 
		\hline
		$\textit{dD}$ & 0.317$\pm$ 0.024 & 31.5$\%$ & 0.2522$\pm$0.028 & -0.27$\%$ &0.33$\%$ &-0.3$\%$ \\ 
		\hline
		$\textit{vD}$ & 0.277$\pm$ 0.024 & 14.9$\%$ & 0.2523$\pm$0.028 & -0.23$\%$ &0.21$\%$ &- \\ 
		\hline
	\end{tabular}}
	\begin{tablenotes}
		\item Table 1. Measured intensities of signal and decoy pulses for three types of predecessors. 
		The Patterning effect is defined as the deviation from the reference patterns ($sS$ and $sD$). \textcolor{black}{The data for MMZI \cite{lu2021} and Sagnac \cite{roberts2018} modulators are listed for comparison.} 
	\end{tablenotes}
\end{center}

\textcolor{black}{As compared in Table~1, the IQ modulator has similar performance as the MMZI~\cite{lu2021} and is arguably better than the Sagnac modulator \cite{roberts2018}.} However, for the IQ modulator, the decoy to signal intensity ratio is 1:4 and is not adjustable. It is therefore necessary to evaluate how such inflexibility affect QKD's performance.
We simulated the impact of intensity ratio on the secret key rate \cite{RN99}. Here, we use a system clock frequency of 1.25 GHz, $\mu_{S} = 0.45$ and an encoding probability of signal, decoy and vacuum states of $6:1:1$.
\begin{figure}[ht]
	\centering\includegraphics[width=\columnwidth]{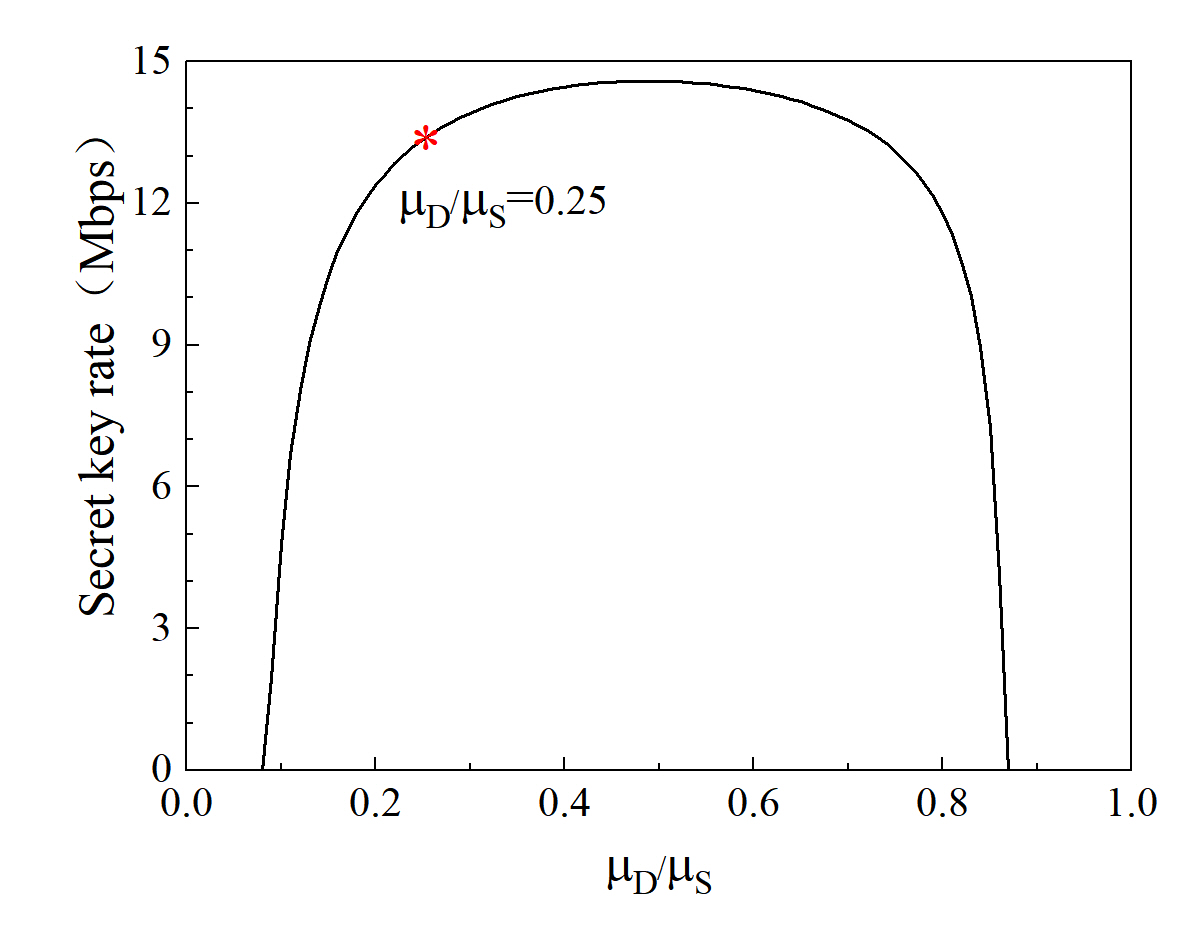}
	\caption{The \textcolor{black}{simulated} dependence of secret key rate on the ratio of decoy to signal state intensity for a 1.25~GHz decoy-state QKD system \textcolor{black}{over a 10~km (2~dB) fiber channel} using the Lim14 procotol \cite{RN99}.}
\end{figure}
\textcolor{black}{For detector performance, we adopt parameters from a home-designed avalanche photodiode with ultra-narrowband interference circuit\cite{Fan}}. The detection efficiency is 0.21, the dark count rate is $5.4\times 10^{-7}$, \textcolor{black}{the afterpulsing probability is 0.01}, and a transmission distance is 10~km \textcolor{black}{(2~dB)}.  The receiver's interferometer loss is 2dB and an error correction efficiency of 1.05~\cite{mao22}. We also considered the finite size effect and the post-processing block size is 100Mb\cite{RN65}. 
Assuming an adjustable modulator, we compute the secure key rate as a function of the ratio of decoy to signal intensity, $\mu_D/\mu_S$. Figure~4 shows the calculation result. 
The system can achieve a maximum key rate (SKR) of  14.59~Mb/s when the ratio is 0.49. When the intensity ratio changing from 0.2 and 0.8, the secret key rate does not change significantly. For IQ modulator($\mu_D/\mu_S = 0.25$), the SKR is reduced by 8.5\% to 13.34~Mb/s .
Therefore, even through we cannot use an IQ modulator to produce decoy states of an arbitrary intensity, it will not result in great reduction to the secret rate of a decoy-state QKD system.  

\section{Conclusion}
Here, for the \textcolor{black}{patterning} effect of the intensity modulator in the application process of high-speed QKD system, we analyze the transfer functions of IQ modulator. Through the transfer function, we demonstrate the advantages of the IQ modulator, and analyze how to use the IQ modulation to prepare three stable intensity states.
Through experimental verification, we confirm that the IQ modulator can effectively suppress the \textcolor{black}{patterning} effect while ensuring that the output intensity in the vacuum state always maintain perfect destructive interference (extinction ratio: 31 dB). Finally, we proved through theoretical simulation that even though the IQ modulator cannot adjust the ratio of decoy state and signal state, it has no prohibitive influence on the secret key rate.
The experimental result indicates that the IQ modulator is suitable 
for future trends and will help the decoy state method to be flawlessly applied to QKD systems in practice.
\begin{backmatter}
\bmsection{Funding}
    National Natural Science Foundation of China (\textcolor{black}{62250710162})
\bmsection{Disclosures}
    The authors declare no conflicts of interest.
\bmsection{Data Availability Statement}
    Data underlying the results presented in this paper are not publicly available at this time but may be obtained from the authors upon reasonable request.
\end{backmatter}

\bibliography{DecoyMod}
\bibliographyfullrefs{DecoyMod}

%\appendix*
%\input{sections/appendix1.tex}

%\section{References}

%\bigskip
%\noindent Add citations manually or use BibTeX. See \cite{Zhang:14,OPTICA,FORSTER2007,testthesis,manga_rao_single_2007}.

% Bibliography
%\bibliography{DecoyMod}

% Full bibliography added automatically for Optics Letters submissions; the following line will simply be ignored if submitting to other journals.
% Note that this extra page will not count against page length
%\bibliographyfullrefs{DecoyMod}

%Manual citation list
%\begin{thebibliography}{1}
%\bibitem{Zhang:14}
%Y.~Zhang, S.~Qiao, L.~Sun, Q.~W. Shi, W.~Huang, %L.~Li, and Z.~Yang,
 % \enquote{Photoinduced active terahertz metamaterials with nanostructured
  %vanadium dioxide film deposited by sol-gel method,} Opt. Express \textbf{22},
  %11070--11078 (2014).
%\end{thebibliography}

% Please include bios and photos of all authors for aop articles
\ifthenelse{\equal{\journalref}{aop}}{%
\section*{Author Biographies}
\begingroup
\setlength\intextsep{0pt}
\begin{minipage}[t][6.3cm][t]{1.0\textwidth} % Adjust height [6.3cm] as required for separation of bio photos.
  \begin{wrapfigure}{L}{0.25\textwidth}
    \includegraphics[width=0.25\textwidth]{john_smith.eps}
  \end{wrapfigure}
  \noindent
  {\bfseries John Smith} received his BSc (Mathematics) in 2000 from The University of Maryland. His research interests include lasers and optics.
\end{minipage}
\begin{minipage}{1.0\textwidth}
  \begin{wrapfigure}{L}{0.25\textwidth}
    \includegraphics[width=0.25\textwidth]{alice_smith.eps}
  \end{wrapfigure}
  \noindent
  {\bfseries Alice Smith} also received her BSc (Mathematics) in 2000 from The University of Maryland. Her research interests also include lasers and optics.
\end{minipage}
\endgroup
}{}

\end{document}